\documentclass[12pt,preprint]{aastex}

\slugcomment{}

\shorttitle{}
\shortauthors{Pipino et al.}

\begin{document}

\title{The outside-in formation of elliptical galaxies}

\author{Antonio Pipino, Francesca Matteucci}
\affil{Dipartimento di Astronomia, Universit\`a di Trieste, Via G.B. Tiepolo 11, 34100 Trieste, Italy}
\email{antonio@ts.astro.it}
\author{and Cristina Chiappini}
\affil{INAF-Osservatorio Astronomico di Trieste, Via G.B. Tiepolo 11, 34100 Trieste, Italy}

\begin{abstract}

In this paper we compare the predictions of a detailed multi-zone chemical
evolution model for elliptical galaxies with the very
recent observations of the galaxy NGC 4697. In particular, the model allows for an initial gas infall and a subsequent galactic wind;
it takes into account detailed nucleosynthesis prescriptions of both type II and Ia supernovae 
and reproduces the main photo-chemical properties of normal ellipticals.
As a consequence of the earlier development of the wind in the outer regions with respect
to the inner ones, we predict an increase of the
mean stellar $[\rm <Mg/Fe>]$
ratio with radius,  in very good agreement with the data for NGC4697.
This finding
strongly supports the proposed outside-in formation scenario for ellipticals.\par
In order to compare theory and observations it is necessary to convert 
either the predicted abundances into line strenght indices or the latter ones into abundances.
We show that, in spite of the good agreement found for the $[\rm <Mg/Fe>]$
ratio, the predicted slope of the mass-weighted
metallicity (Z) gradient does not reproduce the one derived from 
observations, once a calibration to convert 
indices into abundances is applied. This is explained as the consequence of
the different behaviour with metallicity of the
line-strength indices as predicted by a Single Stellar Population (SSP) and
those derived by averaging over a Composite Stellar Population (CSP).  
In fact, in the current literature, the conversion indices-abundances is often
obtained by assuming that ellipticals are SSPs, whereas they are CSPs
at each radius.
In order to better address this issue, we calculate the theoretical ``G-dwarf'' distributions of stars as
functions of both metallicity ([Z/H]) and [Fe/H], showing that they are 
broad and asymmetric that a SSP cannot correctly mimick the mixture of stellar
populations at any given radius. 
We also compute the stellar distribution as a function of the [Mg/Fe] ratio,
which is narrower and with a negligible \emph{skewness} than those as functions of [Z/H] and [Fe/H] and hence can be better
represented by a SSP with an abundance ratio given by the average $[\rm <Mg/Fe>]$ ratio. 
Moreover, we compute the luminosity
distributions of stars for a typical elliptical galaxy as functions 
of [Z/H], [Fe/H] and [Mg/Fe] ratios.
We find that these distributions differ from the ``G-dwarf'' distributions especially at large radii,
except for the one as a function of [Mg/Fe].
Therefore, we conclude that in ellipticals the [Mg/Fe] ratio is the most reliable quantity 
to be compared with observations and is the best estimator 
of the star formation timescale at each radius.
 
\end{abstract}


\section{Introduction}

Metallicity gradients are characteristic of the stellar populations
inside elliptical galaxies. Evidences come from the increase of
line-strength indices (e.g. Carollo et al., 1993; Davies et al., 1993;
Trager et al., 2000) and the reddening of the colours (e.g. Peletier
et al. 1990) towards the centre of the galaxies.  The study of such
gradients provide insights into the mechanism of galaxy
formation, particularly on the duration of the chemical enrichment
process at each radius.  Metallicity indices, in fact, contain
information on the chemical composition and the
age of the single stellar populations (SSP) inhabiting a given
galactic zone.  In particular, by comparing indices related mainly to
Mg (e.g. $Mg_b$, $Mg_2$) to others representative of the Fe abundance
(e.g. $\rm <Fe>$), it is possible to derive the [Mg/Fe] abundance ratio,
which is a very strong constraint for the formation timescale of a
galaxy.  In fact, the common interpretation of the 
$\alpha$-element (O, Mg, Ca,
Si) overabundance relative to Fe, and its decrease with increasing
metallicity in the solar neighbourhood
is due to the different origin of these elements
(time-delay model, Greggio \& Renzini, 1983, Matteucci $\&$ Greggio, 1986), being the former
promptly released by type II supernovae (SNII) and the latter mainly
produced by type Ia supernovae (SNIa) on longer timescales. 
The time-delay model applies also to other objects 
and the [$\alpha$/Fe] versus [Fe/H] relation depends strongly on the star formation history.
For a very short and intense star burst, the [$\alpha$/Fe] ratios
decrease at higher metallicity than in the solar vicinity.
The contrary occurs for a very slow star formation rate
like in irregular galaxies (see Matteucci, 2001).
Pipino \& Matteucci (2004, PM04)
showed that a galaxy formation process in which the most massive
objects form faster and more efficiently than the less massive ones
can explain the photo-chemical properties of ellipticals, in 
particular the increase of [Mg/Fe] ratio in stars with galactic mass 
(e.g. Faber et al., 1992; Carollo et
al., 1993; Davies et al., 1993; Worthey et al., 1992).
Moreover,
from a preliminary analysis of metallicity and colour gradients, PM04
suggested that a single galaxy should form outside-in, namely the
outermost regions form earlier and faster with respect to the central
parts (see also Martinelli et al. 1998).  A natural consequence of this
model and of the time-delay between the production of Fe and that of Mg
is that the mean [Mg/Fe] abundance ratio in the stars should
increase with radius. {\rm We stress that the radial variation
of the [Mg/Fe] abundance ratio has not been previously constrained
in the study presented by PM04.} 
On the other hand, an inside-out formation of ellipticals,
as suggested to explain the abundance gradients in the Milky Way 
(Matteucci \& Fran\c cois 1989, Hou, Prantzos \& Boissier, 2000, Chiappini et al. 2001),
would produce a decrease of the [Mg/Fe] with radius,
since the outermost regions would evolve slower than the inner ones.
Therefore, it is very important to study the gradient of the
[Mg/Fe] ratio, since this can impose strong constraints on the mechanism
of galaxy formation.

In this paper we compare PM04's best model results with the very recent
observations for the galaxy NGC 4697 (Mendez et al. 2005, which
also provide for the first time abundances measured from planetary nebulae (PNe)
in an elliptical galaxy).  The high quality of the optical
spectrum gives us the opportunity to study in great detail the radial
variation of the [Mg/Fe]. 
In order to compare the observations of NGC 4697 with the PM04 model
predictions, we need to adopt a suitable \emph{calibration} relation
to convert our predicted abundances into line-strength indices.
A preliminary study of the effect of the adopted calibration
on the predicted radial gradients has been included in PM04.
Here we present a deeper and self-consistent study on this topic, by
using the same SSPs adopted by Mendez
et al. (2005) to convert their observed indices into abundances.
In particular, we show that the suggested
outside-in formation process is supported
by Mendez et al.'s (2005) data, independently
from the uncertainties in the adopted calibrations.
Moreover, we compare the radial gradients of abundances
in stars as predicted on the basis of
either mass-weighted or luminosity-weighted
averages with observations. We show that the discrepancies arising 
from this comparison should be ascribed to the fact
that a galaxy cannot be described as a unique SSP
and that, at a given galactic radius, several
SSPs contribute to the integrated spectrum.
The paper is organized as follows: in Section 2 we briefly describe
the adopted theoretical model; in Section 3 the predictions
are compared with observations and in Section 4 some conclusions
are drawn.

\section{The model}
The chemical code adopted here is described in full detail 
in PM04, where we address the reader for more details. In particular, being NGC 4697 a $\sim 10^{11}M_{\odot}$ galaxy,
we refer to PM04's Model IIb for the same mass.
This model is characterized by:
Salpeter (1955) IMF, Thielemann et al. (1996) yields for massive stars,
Nomoto et al. (1997) yields for type Ia SNe and 
van den Hoek \& Groenewegen (1997) yields for low-
and intermediate-mass stars (the case with $\eta_{AGB}$ varying with metallicity). 

The model assumes that the galaxy assembles by merging of gaseous
lumps (infall) on a short timescale and suffers a strong star burst
which injects into the interstellar medium a large amount of
energy able to trigger a galactic wind, occurring at different
times at different radii.  After the development of the
wind, the star formation is assumed to stop and the galaxy evolves
passively with continuous mass loss.
We recall here that the assumed star formation efficiency 
is $\nu =10 \rm \, Gyr^{-1}$,
while the infall timescale is $\tau =0.4 \rm \, Gyr$ in the galactic core 
and $\tau =0.01 \rm \, Gyr$
at $1R_{eff}$, respectively. 
These values were chosen by PM04 in order to reproduce the majority of
the chemical and photometric properties of ellipticals. {\rm In particular,
the radial decrease of the infall timescale is needed to reproduce at the same time both
the average metallicity and colour gradients of the observed sample of ellipticals. Only the predictions
on the radial variation of the [Mg/Fe] ratio in stars have not been compared to
the observations. We notice the all the PM04 subcases predict an 
$\alpha$ enhancement which increases with radius, although in different amounts.
This is a consequence of the galactic wind developing earlier where the
potential energy is lower, as already shown by Martinelli et al. (1998)
in models \emph{without infall}. Therefore, it depends only on the \emph{time delay} and
not upon specific model parameters.}
We recall that,
at a given radius, both real and model galaxies are made of a Composite
Stellar Population (CSP), namely a mixture of several SSPs, differing in
age and chemical composition according to the galactic chemical enrichment history,
weighted with the star formation rate.
On the other hand, the line-strength indices are usually tabulated only for
SSPs as functions of their age, metallicity and (possibly) $\alpha$-enhancement.
Therefore, in order to convert the predicted abundances for a CSP into
indices (especially in the case of short burst of star formation), it is typically
assumed that a SSP with a \emph{mean} metallicity is representative of the whole galaxy. 
In other words, we made use of our knowledge about the chemical enrichment
and the star formation histories to calculate either mass- or luminosity-weighted mean
abundance ratios (i.e. $\rm [<\alpha/Fe>],\, [<Fe/H>]\, and\, [<Z/H>]$) for our CSPs, which are then used
to derive the predicted indices for our model galaxies
by selecting a SSP with the same values from the compilations
available in the literature.
In the following we will refer to mass averaged
abundance ratios between the generic elements $\rm X_i$ and $\rm X_j$ with the symbol
$\rm <X_i/X_j>$. Following Pagel \& Patchett (1975, see also Matteucci, 2001):
\begin{equation}
\rm <X_i/X_j>={1\over S_0} \int_0^{S_0} X_i(S)/X_j(S) dS\, ,
\end{equation}
where $\rm S_0$ is the total mass of stars ever born contributing to the 
light at the present time.
The V-luminosity weighted abundances will be denoted
as $\rm <X_i/X_j>_V$. Following Arimoto \& Yoshii (1987), we have:
\begin{equation}
\rm <X_i/X_j>_V=\sum_{k,l}n_{k,l} (X_i/X_j)_l L_{V,k} / \sum_{k,l}n_{k,l} L_{V,k}    \, ,
\end{equation}  
where $n_{k,l}$ is the number of stars binned in the interval
centered around $\rm (X_i/X_j)_l$ with V-band luminosity $\rm L_{V,k}$.
Then we define the $[\rm <X_i/X_j>]$ ratios as $\rm log <X_i/X_j> - log (X_i/X_j)_{\odot}$, taking
the logarithm after the average evaluation (see Gibson, 1996).
Generally the mass averaged {\rm [Fe/H] and [Z/H]} are slightly larger than the luminosity
averaged ones, except for large galaxies (see Yoshii \& Arimoto, 1987, Matteucci et al., 1998). 

\section{Results and discussion}

\subsection{NGC 4697}

\begin{figure}
\plotone{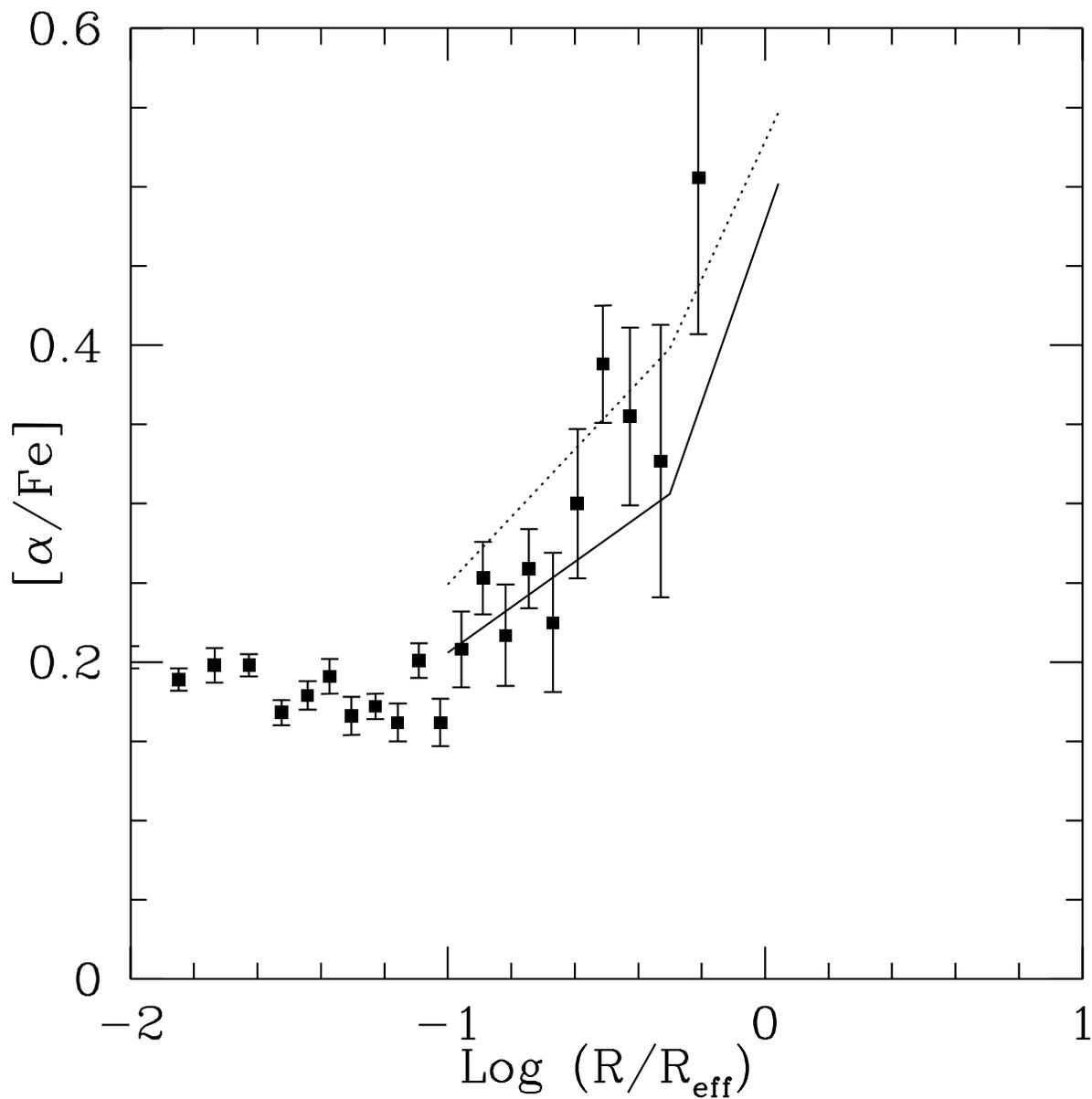}
\caption{PM04's model IIb predictions for the mean mass-weighted [$\rm <\alpha/Fe>$] (solid) and 
luminosity-weighted [$\rm <\alpha/Fe>_V$] (dotted) abundance ratios
in stars as a function of radius compared to the [$\alpha$/Fe] derived for the galaxy NGC 4697 (Mendez et al. 2005, full
squares) by means of TMB SSPs.}
\label{gradaFe}
\end{figure}

\begin{figure}
\plotone{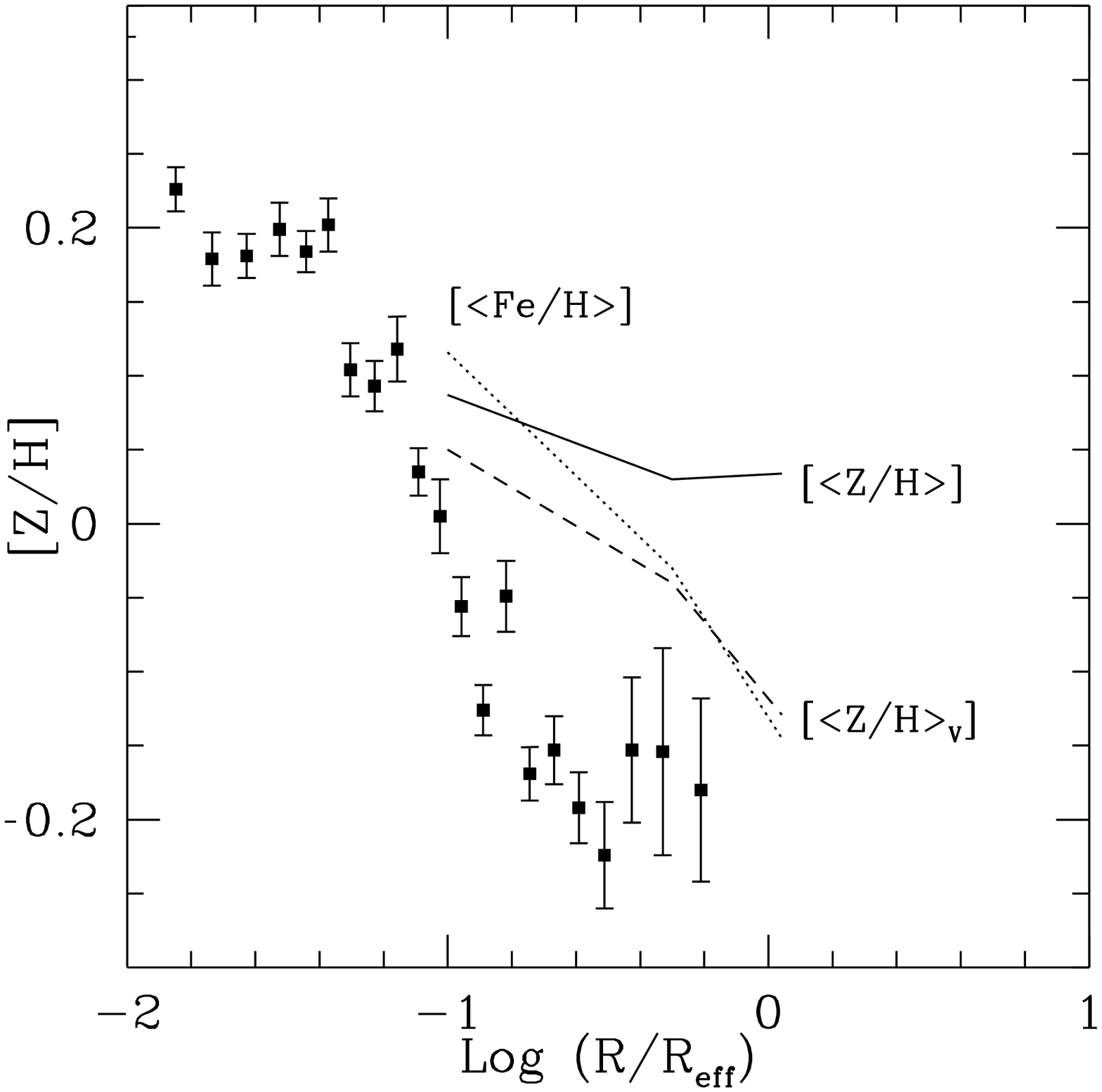}
\caption{PM04's model IIb predictions for the mean mass-weighted [$\rm <Z/H>$] abundance ratio
in stars (solid line) as a function of radius compared to the one derived for the galaxy NGC 4697 (Mendez et al. 2005, full
squares)
by means of TMB SSPs. For comparative purposes we also show the predictions for [$\rm <Fe/H>$] (dotted line)
and [$\rm <Z/H>_V$] (dashed line).}
\label{gradZ}
\end{figure}

In Fig. \ref{gradaFe} we show the predicted radial trends of both the [$\rm <Mg/Fe>$] (solid line) and [$\rm <Mg/Fe>_V$] (dotted line) 
abundance ratios versus the
\emph{observed} one in NGC 4697.  The latter is obtained by Mendez et al. (2005) by converting
the line-strength indices into abundances by means of Thomas et
al. (2003, TMB hereafter) SSPs.  The agreement is remarkable,
especially because we did not tune the input parameters (i.e. 
radius, mass) to exactly match NGC 4697. 
Both the observed increase of
[Mg/Fe] with radius and the slight older age ($\sim$10 Gyr) of the
outermost regions compared to the age of the inner core ($\sim$9 Gyr) confirm
PM04's model predictions, namely an outside-in formation process
in which the central part of the galaxy form stars for
a longer period compared to the most external regions.
This can be explained in terms of galactic winds developing
earlier where the local potential well is shallower (Martinelli et al.
1998).
In Fig. \ref{gradZ} we compare our theoretical gradient for the total
metallicity Z (solid line) with observations. The values for the inner 0.1 $R_{eff}$
are in good agreement with observations, whereas our estimates for the outer regions
are higher than the observed ones. This is because in the conversion
from indices to metallicity it is adopted the approximation of one SSP and a CSP has a higher (mean) metallicity
than a SSP with the same index (Greggio, 1997).
This results from the metallicity distribution of stars and will
be discussed further in the next section.
A steeper slope (remarkably close to the \emph{observed} trend of decreasing
[Z/H] with radius) is predicted in the case of [$\rm <Fe/H>$] (Fig. \ref{gradZ}, dotted line).
A better agreement with observation is also found for the predicted gradient
of [$\rm <Z/H>_V$] (dashed line, see Sec. 3.2).

\begin{figure}
\plotone{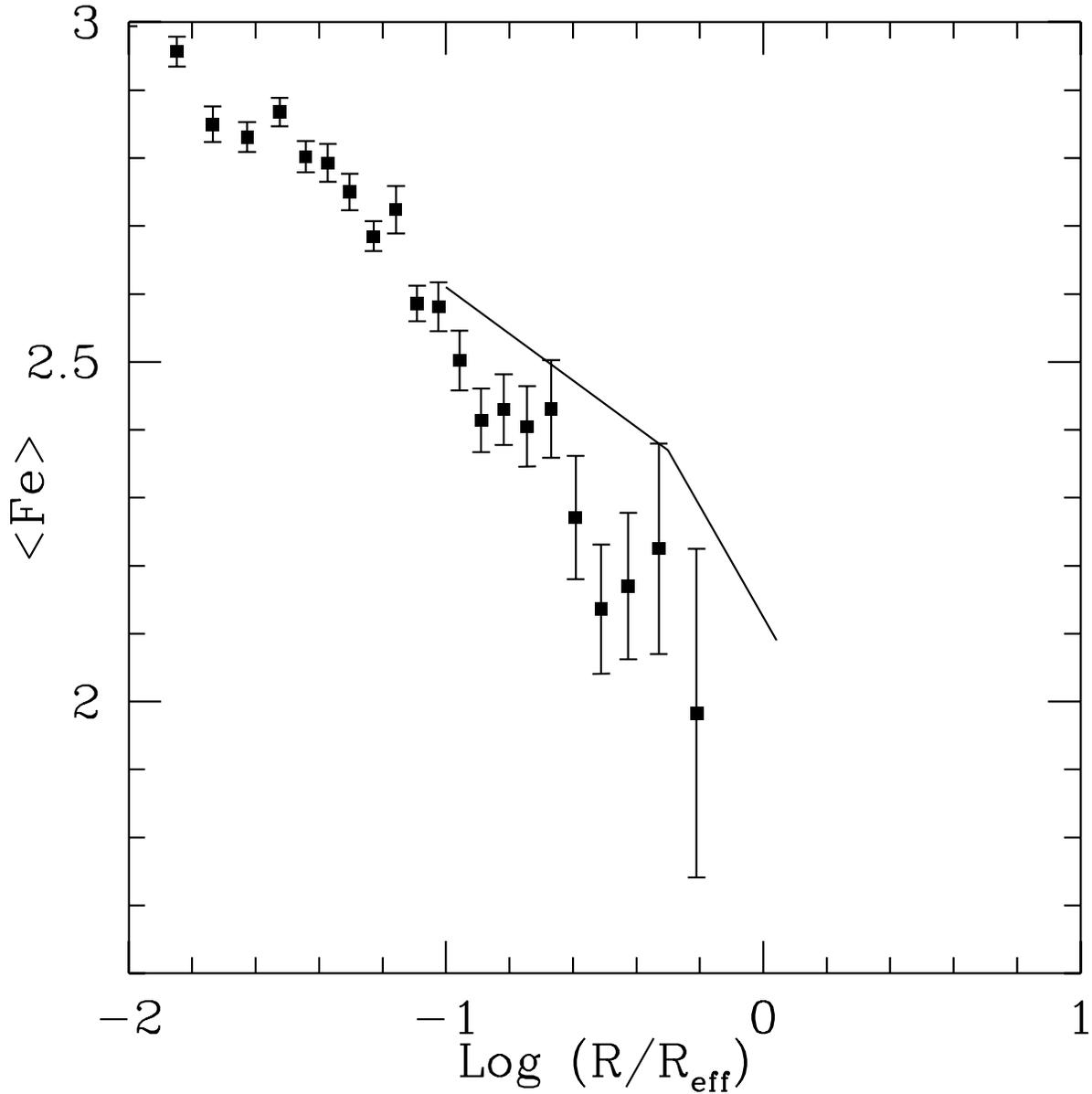}
\caption{PM04's model IIb predictions for the $\rm <Fe>$ line index (obtained
by means of TMB's SSPs from mass-averaged abundances) as a function of radius 
compared to the one observed for the galaxy NGC 4697 (Mendez et al. 2005, full
squares)
.}
\label{grad-fe}
\end{figure}

In order to better understand the origin of this discrepancy, 
here we focus on the procedure typically adopted 
to convert line-strength indices into real abundances.
First of all, we run 
the procedure in the opposite sense, namely we use TMB's SSPs
in order to transform our predicted [$\rm <Z/H>$] and [$\rm <\alpha/Fe>$]
into line-strength indices. The results for $\rm <Fe>$ are shown in
Fig. \ref{grad-fe} (we took also into account the radial age gradient suggested by Mendez et
al. 2005).  We obtain a good agreement for the $\rm <Fe>$ index, whereas the high
predicted [$\rm <Z/H>$] and [$\rm <Mg/Fe>$]  at large radii produce a positive (i.e.
values increasing with radius) gradient
in the $Mg_b$ index, at variance with observations. 
In fact, we obtain $Mg_b = 4.1 $ mag at 0.1
$R_{eff}$, which is in good agreement with the observed value, but at
1 $R_{eff}$ we have $Mg_b = 4.6 \rm mag >> 3.7 \rm mag$ (observed).
This discrepancy was already found by PM04 in the case of  the Matteucci et al. (1998) $\alpha$-enhanced calibration.


The reason for this disagreement is that the [$\rm <Z/H>$] 
(which accounts for the $\sim 60\%$ of the predicted $Mg_b$, according to TMB)
has an almost flat gradient (Fig. \ref{gradZ}), whereas the [$\alpha$/Fe] ratio (which contributes
$\sim 30\%$ to the predicted $Mg_b$) increases with radius (Fig. \ref{gradaFe}). The use of the luminosity-weigthed
values helps in obtaining a negative gradient of $Mg_b$ (namely $Mg_b = 4.1 $ at 0.1
$R_{eff}$ and $Mg_b = 4.0 $ at 1 $R_{eff}$), but the slope
is still too flat with respect to the observations.
As it will be shown in the next section, in fact, the use
of the averages on the luminosity tends to increase the weight
of the low metallicity SSPs with respect to the mass-weighted averages.
On the other hand, we can still
recover the gradient in the Fe indices (Fig. \ref{grad-fe}) because 
it is mainly driven by variations in the $[\rm <Fe/H>]$ alone, for which
we predict a decrease with radius.

\subsection{CSP versus SSP}

Given the discrepancies encountered in the previous section,
we further discuss what is behind the different
behaviours of SSPs and CSPs. The main problem is that, in order
to transform abundances into indices, one has to select a SSP which should be
the closest approximation for the predicted CSP at a given radius. This is usually done by 
assuming that $[\rm <Z/H>]$ and $[\rm <\alpha/Fe>]$ are representative of
the galactic zone and, then, by interpolating among the values tabulated in a suitable 
grid. In the case of TMB's SSP,
the relation [Fe/H]=[Z/H]-0.94[$\alpha$/Fe] has been adopted,
on the basis of a calibration on galactic globular clusters.
This relation can be explained on theoretical grounds (Tantalo \& Chiosi 2004) and
holds also for the abundances in the gas of our model galaxy, but it is not necessarily true for a CSP.
In particular, at 0.1 $R_{eff}$ we predict $[\rm <Fe/H>]=0.116 \ne
[\rm <Z/H>]-0.94[\rm <\alpha/Fe>] = -0.18$, whereas at 1 $R_{eff}$ we predict
$[\rm <Fe/H>]= -0.145 \ne [\rm <Z/H>]-0.94[\rm <\alpha/Fe>] = -0.52$.  
Therefore, since the SSPs are calculated by means of fitting
functions which depend on [Fe/H], this procedure
introduces erroneously low Fe abundances.
This is a general result and depends on the metallicity distribution of the
stellar populations of an elliptical galaxy, as shown by Greggio (1997) for
the indices in the nuclei of ellipticals. In particular, the
calibration of line indices in terms of metallicity
via the comparison with theoretical values for SSPs, is affected
by a systematic effect if a CSP spanning
a substantial metallicity range is present.

\begin{figure}
\epsscale{.90}
\plottwo{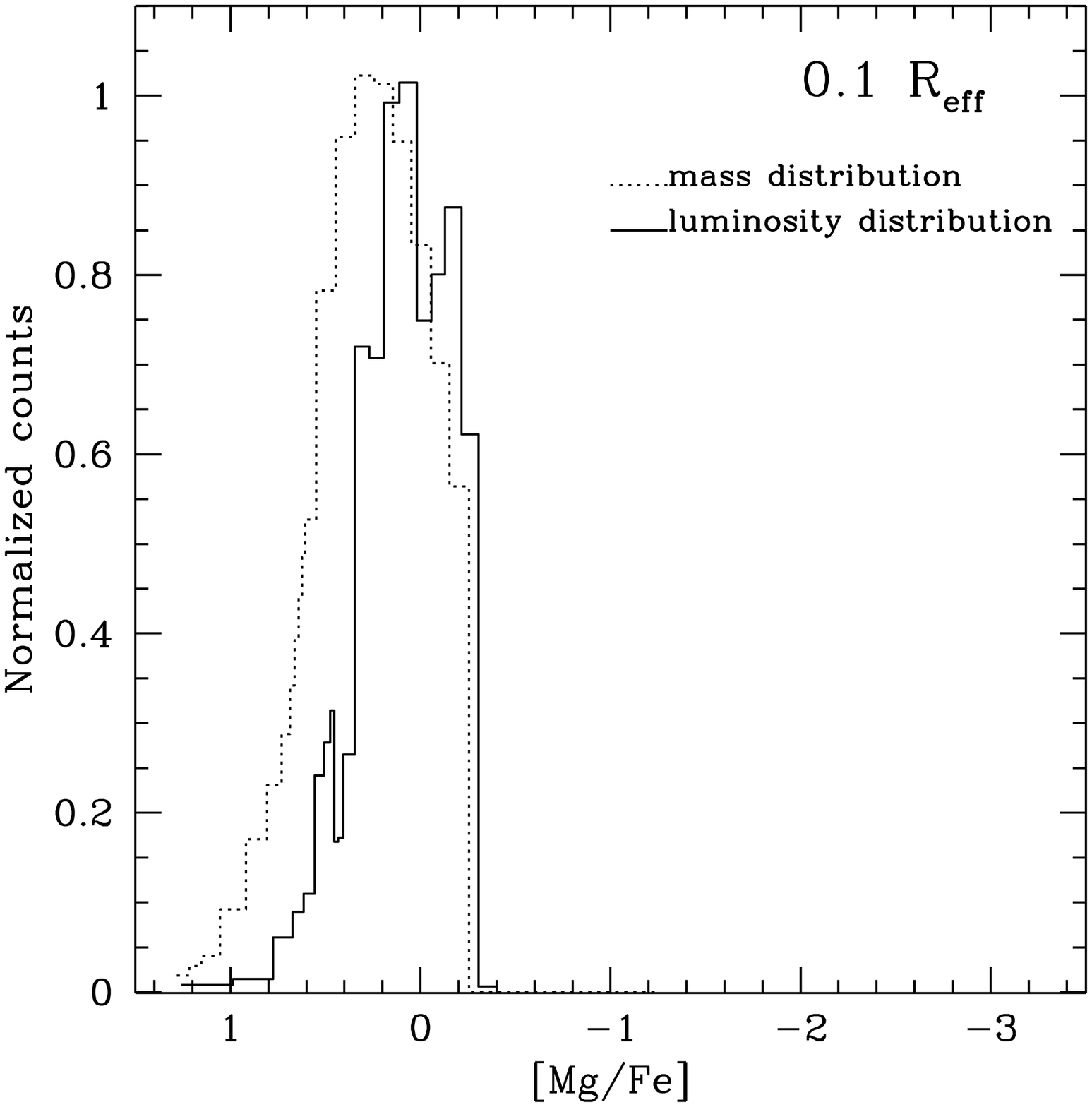}{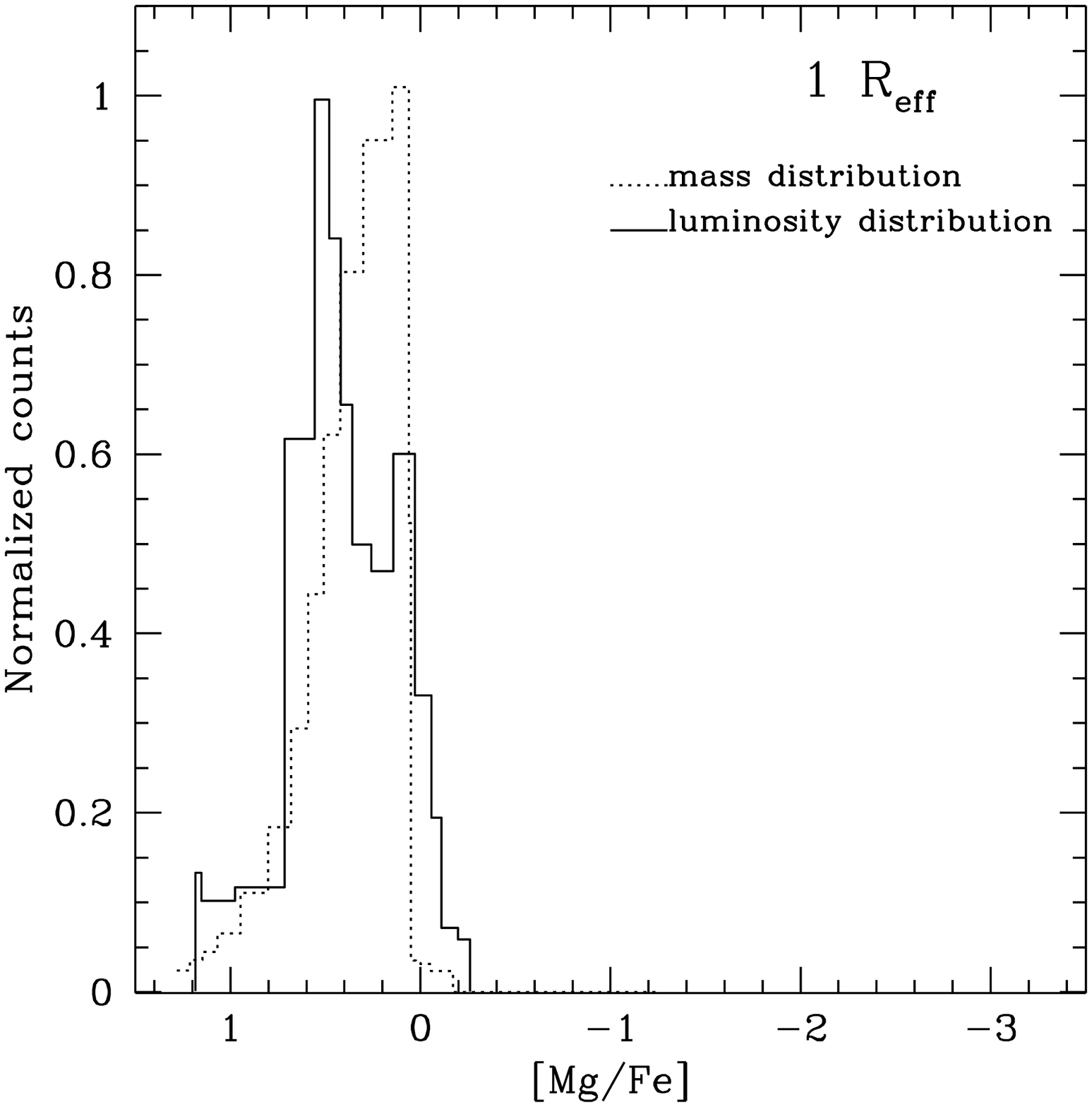}
\plottwo{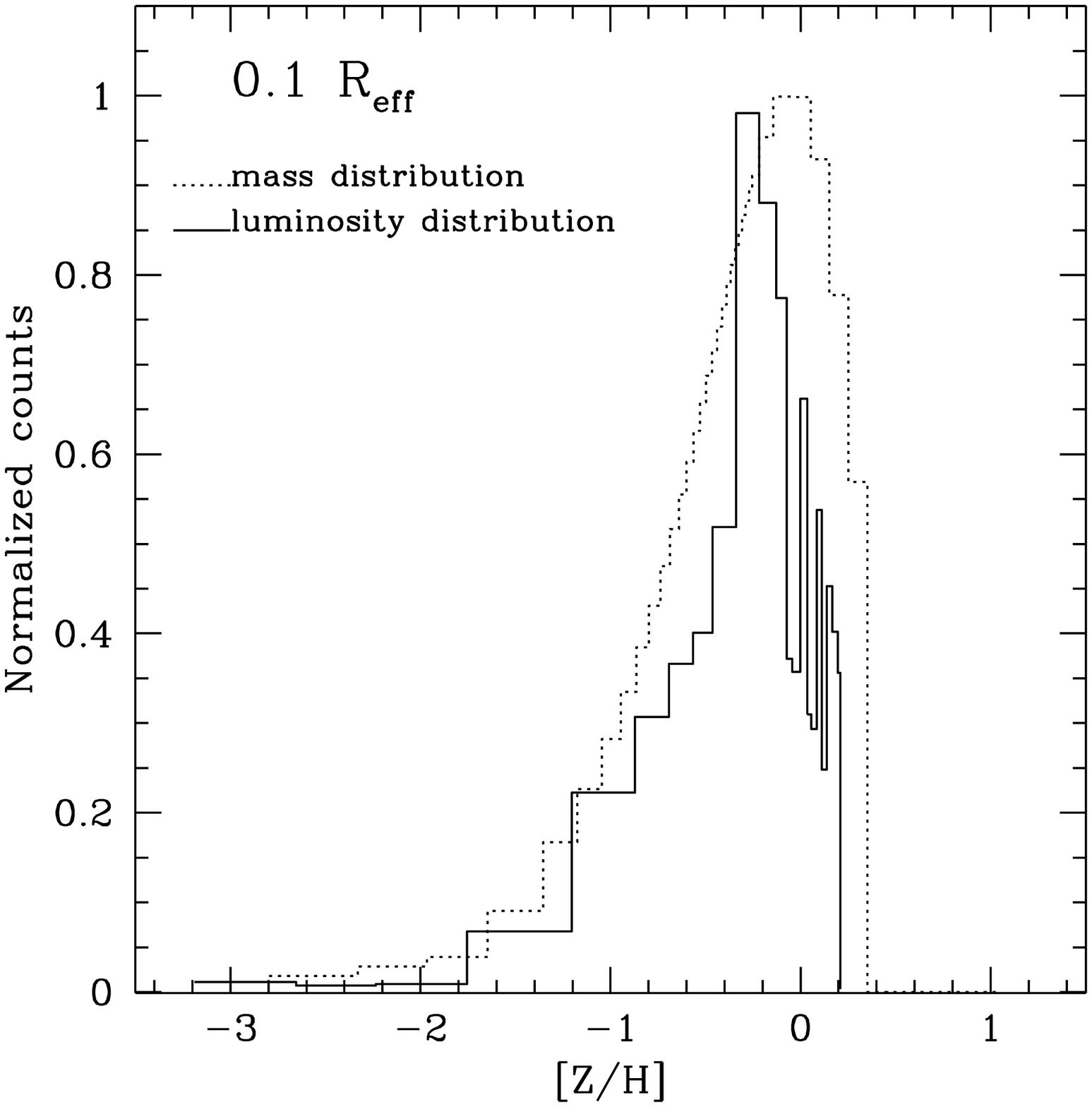}{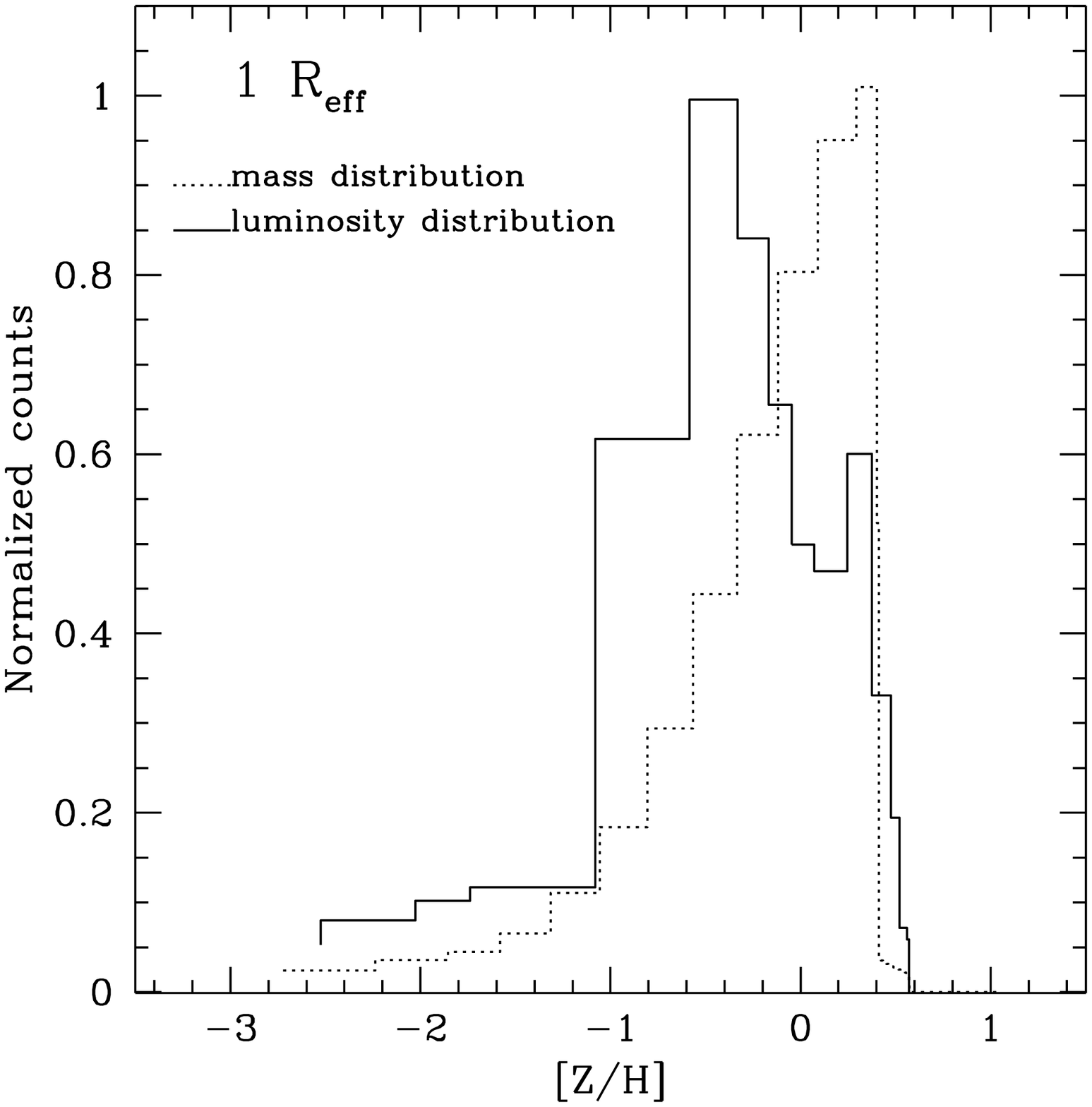}
\plottwo{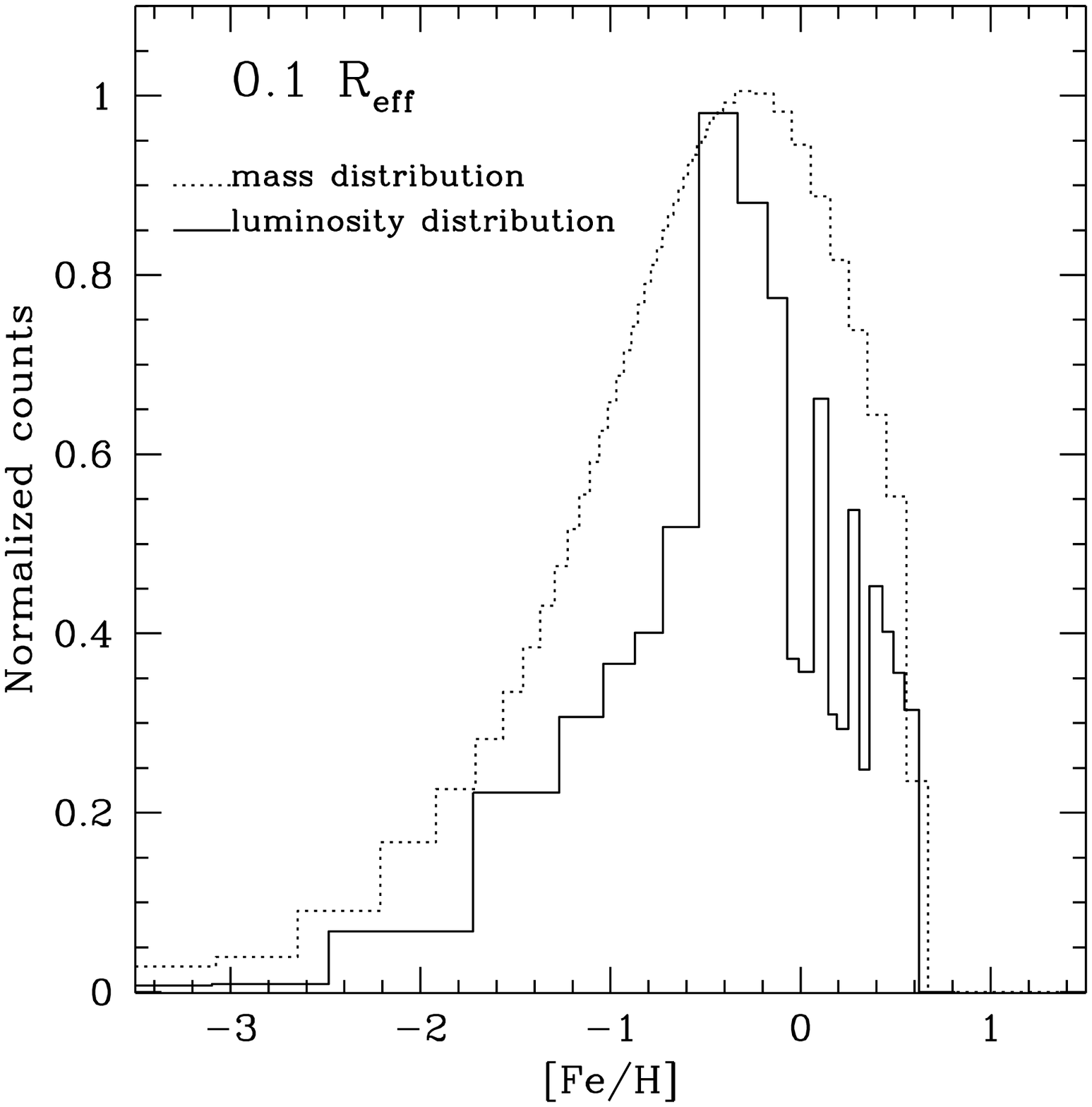}{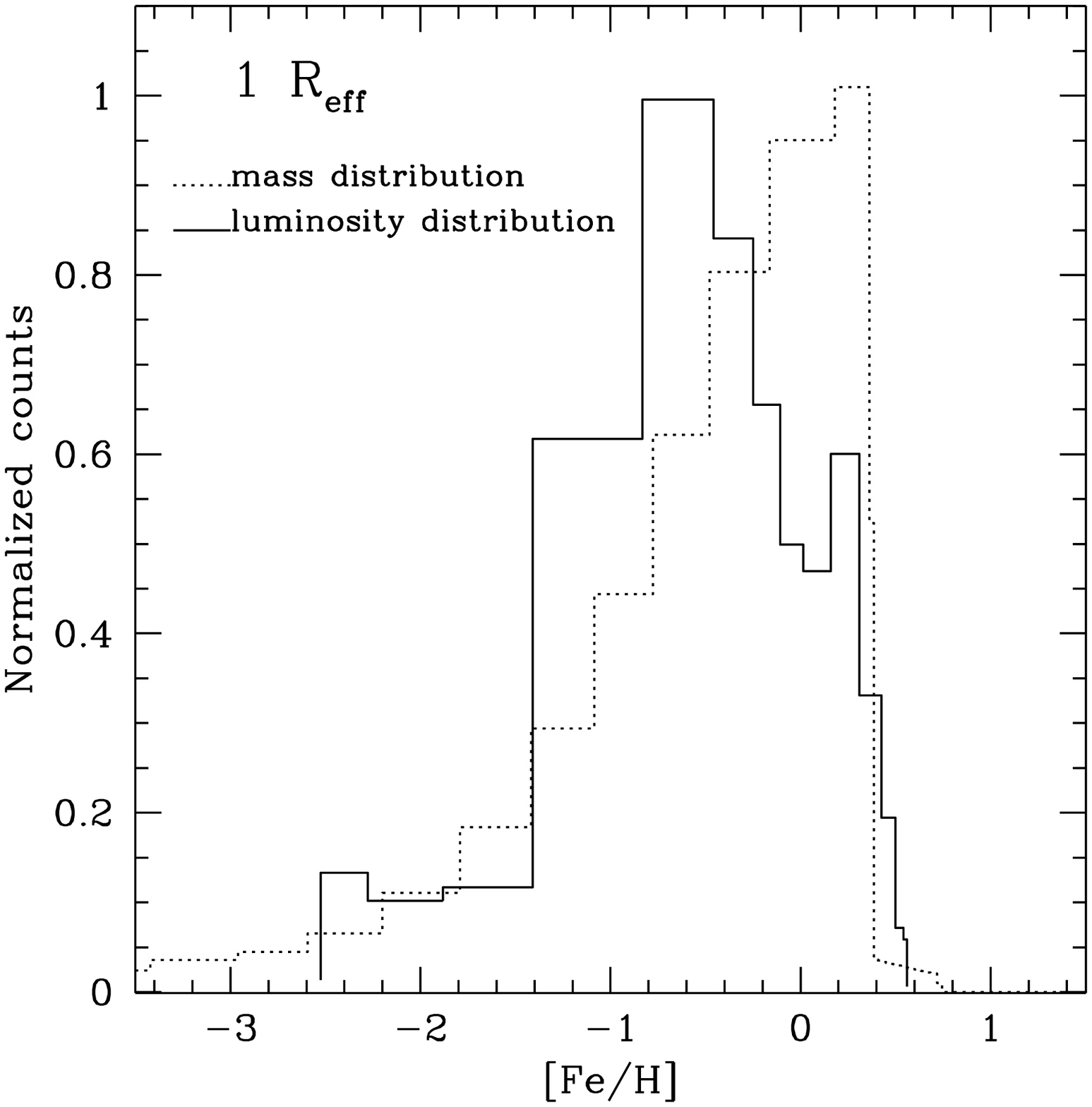}
\caption{``G-dwarf'' distributions for [Mg/Fe] (top row), 
[Z/H] (middle row) and [Fe/H] (bottom row) for the model galaxy
in luminosity (solid line) and mass (dotted line).
\emph{Left column}: values at $0.1 R_{eff}$.
\emph{Right column}: values at $1 R_{eff}$.
The plots are presented in the same scale in order to better
appreciate the differences among the different distributions.
}
\label{gdwarf}
\end{figure}

In fact, from Fig. \ref{gdwarf} it is clear that abundance ratios such as
[$\alpha$/Fe] have  narrow and almost symmetric distributions, which
encourages to take the average value as representative of the whole
CSP. The robustness of the [$\alpha$/Fe] ratios as constraints for
the galactic formation history is testified by the fact that 
$[\rm <\alpha/Fe>]\simeq [\rm <\alpha/Fe>_V]$, having very similar
distributions.
On the other hand, the ``G-dwarf'' plots for [Z/H] and [Fe/H]
exhibit a higher degree of asymmetry and have a broader distribution
with respect to the [$\alpha$/Fe] distribution.
{\rm In particular, if we measure the asymmetry by means of the distribution \emph{skewness}
\footnote{Given a set of $N$ measured values $x_j$ with a mean $\mu$, the skewness
is  $Skew = {1\over N} \sum_{j=1}^N ({x_j - \mu \over \sigma})^3$\, ; where
$\sigma$ is the distribution's standard deviation (Press et al., 1986).},
we find that this parameter is much larger for the [Z/H] and [Fe/H]
distributions than for the case of the [$\alpha$/Fe] one by
more than one order of magnitude. Moreover, the asymmetry
increases going to large radii, especially for the [Z/H]
distribution, in which the skewness increases by a factor
of $\sim$7 with respect to the inner regions. Therefore, it is not surprising that}
the $[\rm <Z/H>]$ value does not
represent the galaxy at large radii,
and hence, we stress that care should be taken when
one wants to infer the real abundances of the stellar components
for a galaxy by comparing the observed indices (related
to a CSP) with the theoretical ones (predicted
for a SSP). Only the comparison based on the $[\rm <\alpha/Fe>]$ ratios
seems to be robust.

Another possible source of discrepancies is the fact
that luminosity-weighted averages (which are more closely related
to the observed indices) and mass-weighted averages (which
represent the real distributions of the chemical
elements in the stellar populations) might differ more in the most external
zones of the galaxy.

This was already shown in Fig.
\ref{gradZ} (dashed line), where the radial gradient of $[\rm <Z/H>_V]$
is in better agreement with the observed slope.
In fact, in the luminosity-weighted averages,
the stars with lower metallicity have a lower mass-to-light ratio in the V-band
with respect to the high [Z/H] ones;
therefore, the luminosity-weighted averages shift towards systematically lower values
than the mass-weighted ones and the difference between the true
[Fe/H] and the one obtained from the TMB calibration decreases.
This is particularly true in the most external region (see Fig. \ref{gdwarf}) ,
at variance with the typical assumption that
$[\rm <Z/H>]\simeq [\rm <Z/H>_V]$ for sufficiently old
systems, which is still valid, however, for the galaxy as a whole (Matteucci et al., 1998, Gibson, 1996, see 
Sec. 2) and for the inner regions.

Beyond one effective radius,
PM04's best model predicts that both star formation
and infall timescales are so short that a flattening
of the radial gradients is expected.
This result seems to be in agreement with the metallicity gradient
inferred from PNe by Mendez et al. (2005), although
they provide only lower limits.
We hope that in the future more data from PNe {\rm (or other objects
as globular clusters)} will be available
in order to better constrain the process of galaxy formation \rm
by means of new and independent tools.
This could be a first step towards a better knowledge
of the stellar abundances,
as it happens for the stars in the Milky Way.
\rm
In fact, the use of observations coming from single
objects will avoid the use of the calibration relations
with all the uncertainties linked to the population synthesis modelling.

\rm 
We conclude by stressing that NGC4697 is a well-observed single object, whose properties
have been compared to a model created to reproduce
the average properties of ellipticals. We find that the agreement is remarkable.
We notice that, recently, an increasing radial
gradient for the [O/Fe] ratio has been observed by Proctor et al. (2005) in the
galaxy NGC 821. The inferred radial trend is very similar to the one
observed in NGC 4697 (see their Fig. 8).
We avoided to show NGC 821's observations, in order not
to mix data obtained with different quality and calibrations.
A larger compilation of [$\alpha$/Fe] gradients is the one by
Mehlert et al. (2003). They suggest that, at least
for ellipticals in the Coma cluster, the average [Mg/Fe] gradient
is zero (although objects with gradients as high as in NGC 4697
are also present).
On the other hand, Ogando et al. (2005) presented
observed massive ellipticals with strong gradients
(up to $\Delta [Z/H] / \Delta log (R/R_{eff}) = -1$) increasing
with galactic mass. These slope are steeper than the typical
values found by Mehlert et al. (2003) in their sample.
For these galaxies we expect a very strong [$\alpha$/Fe] gradient,
but this should still be compared by observations.
Massive galaxies which exhibit shallower gradients, instead,
could have a prolonged formation history, perhaps affected
by mergers (e.g. Kobayashi, 2004).

Given the importance of the subject in the study of galaxy formation, 
and the issues arising when SSPs are taken as representative of CSPs,
we hope that wider, more accurate and homogeneous
samples of [Mg/Fe] gradients will be the goal of future
observational campaigns.

Finally, we notice that radial variations in the $\alpha$ enhancement alone
can be reproduced in several ways. For example, a constant [$<Mg/Fe>$]
along the galactic radius can be a result of either a formation process which has a 
timescale independent from the galactocentric distance 
or by a radial change of the IMF (see e.g. Martinelli et al. 1998).
In any case, it is the whole picture coming from the largest possible set of 
spectro-photometric observables which helps is determining the
most likely formation scenario.
\rm

\section{Conclusions}
By means of the comparison between PM04's best model predictions and
the recent and high quality data on NGC 4697 (Mendez et al., 2005),
we are able to derive some conclusions on the radial trend of abundances and abundance
ratios in the stellar component of elliptical galaxies.
Our main conclusions are:

\begin{itemize}
\item PM04's best model prediction of increasing [$\rm <\alpha/Fe>$] ratio
with radius is in very good agreement with the observed
gradient in [$\alpha$/Fe] of NGC 4697.
This strongly suggests an outside-in galaxy formation scenario
for elliptical galaxies that show strong gradients.
\item By comparing the radial trend of [$\rm <Z/H>$] with the 
\emph{observed} one,
we notice a discrepancy which is due to the fact that a
CSP behaves in a different way with respect to a SSP.
In particular the predicted gradient of [$\rm <Z/H>$] is flatter
than the observed one at large radii.
Therefore, this should be taken into account when 
estimates for the metallicity of a galaxy are derived
from the simple comparison between the observed line-strength index
and the prediction for a SSP, a method currently adopted in the literature.
\item Luminosity-weighted abundances might differ from mass-averaged ones
at large radii even for old (i.e. $\sim 10$ Gyr old) stellar populations,
at variance with what is commonly assumed.
\item Abundance ratios such as [Mg/Fe] are less affected by the discrepancy
between the SSPs and a CSP, since
their distribution functions are narrower and more symmetric.
Therefore, we stress the importance of such ratio as the
most robust tool to estimate the duration of the galaxy
formation process.
\end{itemize}

\acknowledgments
This work received the support from
MIUR under COFIN03 prot. 2003028039.
Useful discussions with I.J. Danziger are acknowledged.
We thank the referee for useful comments.

\clearpage


\begin{thebibliography}{}
\small
\bibitem []{}Arimoto, N., $\&$ Yoshii, Y. 1987, A$\&$A, 173, 23 
\bibitem []{}Carollo, C.M., Danziger, I.J., $\&$ Buson, L. 1993, MNRAS, 265, 553
\bibitem[]{} Chiappini, C., Matteucci, F., \& Romano, D.\ 2001, \apj, 554, 1044
\bibitem []{}Davies, R.L., Sadler, E.M., $\&$ Peletier, R.F., 1993, MNRAS, 262, 650
\bibitem []{}Faber, S.M., Worthey, G., $\&$ Gonzalez, J.J. 1992, in IAU Symp. n.149,
eds. B. Barbuy $\&$ A. Renzini, p. 255
\bibitem []{}Gibson, B.K., 1996, MNRAS, 278, 829
\bibitem []{}Greggio, L., 1997, MNRAS, 285, 151
\bibitem []{}Greggio, L., Renzini, A., 1983 in 
``Frascati Workshop on First Stellar Generations'',  Memorie Societa Astronomica Italiana, vol. 54, 
p. 311
\bibitem []{}Hou, J., Prantzos, N., Boissier, S., 2000, A\&A, 362, 921
\bibitem []{}Jimenez, R., Padoan, P., Matteucci, F., Heavens, A.F., 1998, MNRAS, 299, 123
\bibitem []{}Kobayashi, C., 2004, MNRAS, 347, 740
\bibitem []{}Martinelli, A., Matteucci, F., Colafrancesco, S., 1998, MNRAS, 298, 42
\bibitem []{}Matteucci, F. 1994, A$\&$A, 288, 57
\bibitem []{}Matteucci, F. 2001, The chemical evolution of the Galaxy, Kluwer Academic Publishers, Dordrecht
\bibitem []{}Matteucci, F., Fran\c cois, P., 1989, MNRAS, 239, 885
\bibitem []{}Matteucci, F., Ponzone, R., Gibson, B.K., 1998, A$\&$A, 335, 855
\bibitem []{}Mendez, R.H., Thomas, D., Saglia, R.P., Maraston, C., Kudritzki, R.P., \& Bender, R., 2005,
ApJ, 627, 767
\bibitem []{}Nomoto, K., Hashimoto, M., Tsujimoto, T., Thielemann, F.K., Kishimoto, 
N., Kubo, Y., Nakasato, N., 1997, Nuclear Physics A, A621, 467
\bibitem []{}Ogando, R.L.C., Maia, M.A.G., Chiappini, C., Pellegrini, P.S.,
Schiavon, R.P., da Costa, L.N., 2005 (astro-ph/0509142) 
\bibitem []{}Pagel, B.E.J., $\&$ Patchett, B.E. 1975, MNRAS, 172, 13 
\bibitem []{}Peletier, R.F., Davies, R.L., Illingworth, G.D., Davis, L.E., Cawson, M. 1990, AJ, 100, 1091
\bibitem []{}Pipino,  A.,  Matteucci,  F. 2004, MNRAS, 347, 968 (PM04)
\bibitem []{}Prantzos, N., \& Boissier, S., 2000, MNRAS, 313, 338
\bibitem []{}Press  W.H., Flannery  B.P., Teukolsky  S.A., Vetterling  W.T., 1986, Numerical Recipes,
Cambridge Univ. Press
\bibitem []{}Proctor, R.N., Forbes, D.A., Forestell, A, \& Gebhardt, K., (astro-ph/0506523)
\bibitem []{}Salpeter, E.E., 1955, ApJ, 121, 161
\bibitem []{}Tantalo, R., \& Chiosi, C., 2004, MNRAS, 353, 917
\bibitem []{}Thielemann, F.K., Nomoto, K., Hashimoto, M. 1996, ApJ, 460, 408 
\bibitem []{}Thomas, D., Maraston, C., $\&$ Bender, R., 2003, MNRAS, 339, 897 
\bibitem []{}Trager, S.C., Faber, S.M., Worthey, G., Gonzalez, J.J., 2000a, AJ, 119, 1654
\bibitem []{}van den Hoek, L.B., Groenewegen, M.A.T. 1997, A$\&$AS, 123, 305
\bibitem []{}Yoshii, Y., \& Arimoto, N., 1987, A$\&$A, 188, 13 
\bibitem []{}Worthey, G., Faber, S.M., $\&$ Gonzalez, J.J. 1992, ApJ, 398, 69

\end{thebibliography}
\end{document}